\documentclass[aps,prb,superscriptaddress,showpacs,preprint,floatfix]{revtex4-1}

\usepackage{epsfig,amsmath,amssymb,txfonts}

\newcommand*{\abinit}[0]{\textit{ab~initio}}
\newcommand*{\Abinit}[0]{\textit{Ab~initio}}

\newcommand*{\et}[0]{\textit{et~al.}}

\newcommand*{\eps}[0]{\varepsilon}
\newcommand*{\x}[0]{\times}
\newcommand*{\EE}[1]{\times 10^{{#1}}}

\newcommand*{\kB}[0]{k_{\text{B}}}
\newcommand*{\eV}[0]{\text{eV}}
\newcommand*{\meV}[0]{\text{meV}}
\newcommand*{\DR}[0]{\underline{D}}
\newcommand*{\Dk}[0]{\widetilde{D}}

\newcommand*{\qv}[0]{\vec q}

\newcommand*{\Rv}[0]{\vec R}
\newcommand*{\rv}[0]{\vec r}

\newcommand*{\uv}[0]{\vec u}
\newcommand*{\uk}[0]{\widetilde{u}}

\newcommand*{\disl}[0]{\rho_{\text{disl}}}
\newcommand*{\rc}[0]{r_\text{c}}
\newcommand*{\DE}[0]{\Delta E}
\newcommand*{\site}[0]{\tilde f}

\newcommand*{\be}[0]{\begin{equation}}
\newcommand*{\ee}[0]{\end{equation}}
\newcommand*{\beu}[0]{\begin{equation*}}
\newcommand*{\eeu}[0]{\end{equation*}}
\newcommand*{\bme}[0]{\begin{multline}}
\newcommand*{\eme}[0]{\end{multline}}
\newcommand*{\bmeu}[0]{\begin{multline*}}
\newcommand*{\emeu}[0]{\end{multline*}}
\newcommand*{\ba}[0]{\begin{array}}
\newcommand*{\ea}[0]{\end{array}}
\newcommand*{\bfig}[0]{\begin{figure}[t]}
\newcommand*{\efig}[0]{\end{figure}}
\newcommand*{\bfigwide}[0]{\begin{figure*}[ht]}
\newcommand*{\efigwide}[0]{\end{figure*}}

\newlength{\wholefigwidth}
\newlength{\smallfigwidth}
\newlength{\halfsmallfigwidth}
\newcommand{\Fig}[1]{Fig.~\ref{fig:#1}}

\newcommand{\Eqn}[1]{Eqn.~\ref{eqn:#1}}
\newcommand{\rcite}[1]{Ref.~\onlinecite{#1}}

\begin{document}

\title{Nanoscale-hydride formation at dislocations in palladium: \Abinit\ theory and incoherent inelastic neutron scattering measurements}

\author{Dallas R. Trinkle}
\email{dtrinkle@illinois.edu}
\affiliation{Department of Materials Science and Engineering, University of Illinois, Urbana-Champaign, Illinois 61801, USA}
\author{Hyunsu Ju}
\author{Brent J. Heuser}
\affiliation{Department of Nuclear, Plasma, and Radiological Engineering, University of Illinois, Urbana-Champaign, Illinois 61801, USA}
\author{Terrence J. Udovic}
\affiliation{NIST Center for Neutron Research, National Institute of Standards and Technology, Gaithersburg, Maryland 20899, USA}

\date{\today}
\begin{abstract}
Hydrogen arranges at dislocations in palladium to form nanoscale hydrides, changing the vibrational spectra.  An \abinit\ hydrogen potential energy model versus Pd neighbor distances allows us to predict the vibrational excitations for H from absolute zero up to room temperature adjacent to a partial dislocation and with strain.  Using the equilibrium distribution of hydrogen with temperature, we predict excitation spectra to explain new incoherent inelastic neutron-scattering measurements.  At 0K, dislocation cores trap H to form nanometer-sized hydrides, while increased temperature dissolves the hydrides and disperses H throughout bulk Pd.
\end{abstract}

\maketitle

\section{Introduction}
The increasing needs for renewable energy---and issues of production, storage and transportation of energy---motivates interest in hydrogen for energy storage.\cite{Schlapbach2001}  At a fundamental level, open questions remain about how hydrogen acts in metals, despite a long legacy of study.\cite{Myers1992, Pundt2006} Palladium is an ideal metal to study hydrogen behavior due to the strong catalytic behavior of the Pd surface facilitating hydrogen adsorption, favorable $T-p_{\text{H}_2}$ thermodynamic properties, and that hydrogen acts as an ideal lattice gas in Pd.\cite{Fukai1993}  Neutron-scattering characterization is useful,\cite{Fukai1993} in part to a scattering interaction mediated by neutron-nuclear properties and available incident neutron energies similar to those associated with lattice vibrations.  Coherent inelastic neutron scattering gave the first phonon dispersion measurement of a metal hydride (Pd-H and -D).\cite{Rowe1974}  The hydrogen-dislocation trapping interaction in Pd has remained of significant interest over the last four decades\cite{Flanagan1976, Pundt2006} because of the favorable Pd-H properties mentioned above and that Pd can be heavily deformed by hydride cycling across the miscibility gap.\cite{Heuser1991}

Mobile solutes---substitutional and interstitial---arrange themselves in a crystal to minimize the free energy; with non-uniform strains, the arrangement reflects the energy changes from strain.  For an edge dislocation, compressive and tensile strains produce areas that are depleted and enhanced with solute concentration---a ``Cottrell atmosphere.''\cite{Fiore1968}  Cottrell atmospheres produce time-dependent strengthening mechanisms like strain-aging in steels and the Portevin-Le~Chatelier effect in aluminum alloys,\cite{Dieter1986} and the rearrangement of hydrogen from dislocation strain fields affects dislocation interactions.\cite{Ferreira1998}  The dislocation core---where the continuum description of the strain fields breaks down---provides the largest distortions in geometry and the attraction of solutes to this region is crucial for solute effects on strength.\cite{TrinkleSSS2005, Curtin2006, Yasi2010}  Tensile strain also lowers the vibrational excitation for H, and, in a dislocation core, broken symmetry splits the excitations.\cite{Lawler2010}  The vibration of Pd next to H changes the local potential energy for each H atom, broadening the vibrational excitations.  Additionally, the vibrational excitations of the light hydrogen atom are significantly changed by anharmonicity.\cite{Elsasser1991,Krimmel1994}  We treat all of these effects: non-uniform hydrogen site occupancy due to strain and H-H interaction, quantum-thermal vibrational displacements for neighboring Pd, and the anharmonic potential energy to determine the causes of changes to the vibrational spectra with temperature.  Experimentally, in situ inelastic neutron scattering averages over different H sites to give a direct measurement of H environment.  We compare our \abinit\ treatment of hydrogen sites and anharmonic vibrational excitations with incoherent inelastic neutron-scattering measurements to observe the formation and dissolution of nanoscale hydrides around dislocation cores in palladium.

\section{Methods}
Incoherent inelastic neutron scattering (IINS) using the Filter Analyzer Neutron Spectrometer (FANS) at the NIST Center for Neutron Research\cite{Udovic2008} measure the vibrational density of states of trapped hydrogen in polycrystalline Pd as a function of temperature.  FANS scans the incident neutron energy and records the intensity that passes through a Be-Bi-graphite composite neutron filter.  Sample preparation procedures and material are identical to Heuser~\et,\cite{Heuser2008} with $\sim$100 grams of polycrystalline Pd sheet measured at 4K, 100K, 200K, and 300K.  Palladium sheet supplied by Alpha Aesar was cold-rolled in the as-received condition, and further deformed by cycling twice across the hydride miscibility gap.\cite{Heuser1991}  It was held under vacuum at room temperature for several days and then annealed for 8 hours at $\sim$400K to completely outgas the sample.  The subsequent measured pressure reduction in a closed volume at room temperature using a portable hydrogen gas loading apparatus gives a total hydrogen concentration of 0.0013 [H]/[Pd], corresponding to a total hydrogen inventory of 1.3 mg.  The IINS measurements were performed in an Al measurement can sealed with indium wire.\cite{Heuser2008}  This can was isolated with an all metal vacuum valve, mounted to the FANS instrument, and cooled to 4K.  Subsequent measurements were performed at 100K, 200K, and 300K.  The sample was then outgassed at $\sim$420K for $\sim$48 hours completely remove all hydrogen.  The zero-concentration background was measured from the out-gassed sample in the Al can at 4K, 100K, 200K, and 300K.  We also recorded fast neutron background with the sample in place and the detector bank blocked with Cd.\cite{Heuser2008}  The measured hydrogen vibrational density of states in \Fig{DOS-predict} is the normalized net intensity after zero-concentration and fast neutron background subtractions.  In addition, an energy-independent flat background attributed to multi-phonon scattering was subtracted, as discussed in \rcite{Heuser2008}.

Density functional theory calculations for Pd-H\cite{Lawler2010} are performed with \textsc{vasp}\cite{Kresse93,Kresse96b} using a plane-wave basis with the projector augmented-wave (PAW) method\cite{Blochl1994} with potentials generated by Kresse.\cite{Kresse1999}  The local-density approximation as parametrized by Perdew and Zunger\cite{Perdew1981} and a plane-wave kinetic-energy cutoff of 250eV ensures accurate treatment of the potentials.  The PAW potential for Pd treats the $s$- and $d$-states as valence, and the H $s$-state as valence.  The restoring forces for H in Pd change by only 5\%\ compared with a generalized gradient approximation, or including Pd $4p$-states in the valence; our choice of the local-density approximation is computationally efficient, and gives an $\alpha$-Pd lattice constant of 3.8528\AA\ compared with the experimentally measured 3.8718\AA.  To compute the dynamical matrix for Pd, and to relax H at the octahedral site in $\alpha$-Pd, we use a $4\x4\x4$ simple-cubic supercell of 256 atoms, with a $6\x6\x6$ k-point mesh; while the dislocation geometry with 382 atoms uses a $1\x1\x8$ k-point mesh.  For the PdH$_{0.63}$ hydride force-constant calculation, a $3\x3\x3$ simple cubic cell (108 Pd atoms, 68 H atoms) with displacements of 0.01\AA\ for H and Pd atoms and a $8\x8\x8$ k-point mesh.  The electron states are occupied using a Methfessel-Paxton smearing of 0.25eV.  For the H octahedral site in $\alpha$-Pd and the partial dislocation core, atom positions are relaxed using conjugate gradient until the forces are less than 5meV/\AA.

Dislocations produce a distribution of interstitial site strains; to compute the density of strain sites available for hydrogen, we consider a simplified model for the distribution of dislocations throughout the crystal.  We take the dislocation density $\disl$ as given by cylinders of radius $R=1/\sqrt{\pi\disl}$ with an edge dislocation at the center; we assume that the strain in each cylinder is due only to the single edge dislocation at the center.  The volumetric strain $r$ away from the dislocation core and with angle $\theta$ to the slip plane is 
\be
\eps = -\frac{b}{2\pi r}\cdot\frac{1-2\nu}{1-\nu}\sin\theta
= -\sin\theta\cdot\frac{b}{4\pi r}
\ee
for a Poisson's ratio $\nu=1/3$, and where $b=0.298\text{nm}$ is the Pd Burgers vector.  This equation becomes invalid for small $r$; we truncate the expression in the ``core'' of the dislocation.  We can estimate the size of the core by considering the maximum strain of $\pm 5\%$ at the partial core from \rcite{Lawler2010}; then,
\be
\rc = \frac{b}{4\pi(0.05)} = 1.59b \approx \sqrt[4]{6}b
\ee
The line vector of an edge dislocation is $t=\sqrt{6}a_0/2$ with Burgers vector $b=a_0/\sqrt{2}$, and so the core has a volume of $\rc^2 t = 3a_0^3/2 = 6(a_0^3/4)$; hence, there are 6 sites per dislocation line inside this radius.  We assign half the maximum strain of $+5\%$ and half the minimum strain of $-5\%$ corresponding to opposite sides of the partial cores.  Previous \abinit\ calculations of the core give a trapping energy of 0.164eV with a 5\%\ strain;\cite{Lawler2010} the trapping energy matches the decrease in hydrogen energy from a 5\%\ increase in volume---we then model the binding energy for H as linear in the site strain $\eps$: $-0.164\eV(\eps/0.05)$.

With these definitions, we compute the density of strain sites $n(\eps)$ by integrating over our cylinder cross-section from $\rc$ out to $R$.  We consider the 6 core sites (3 attractive and 3 repulsive) separate from this continuum calculation.
\be
\begin{split}
n(\eps) &= \left[\int_0^R d^2r\right]^{-1}\cdot\int_{\rc}^R d^2r\;\delta(\eps - \eps(r,\theta))\\
&=\disl\cdot\int_{\rc}^R rdr\int_0^{2\pi} d\theta\;\delta\left(\eps + \frac{b}{4\pi r}\sin\theta\right)\\
&=2\disl\cdot\int_{\rc}^{\min\{R,b/4\pi\eps\}} rdr \left|\frac{b}{4\pi r}\cos\left(\sin^{-1}\left(\frac{\eps 4\pi r}{b}\right)\right)\right|^{-1}\\
&=2\disl\cdot\int_{\rc}^{\min\{R,b/4\pi\eps\}} dr\frac{r}{\left(\left(\frac{b}{4\pi r}\right)^2 - \eps^2\right)^{1/2}}
\end{split}
\ee
where the delta-function integral is calculated by rewriting the delta function in terms of the two roots $\theta_0 = \sin^{-1}(\eps 4\pi r/b)$.  To simplify the expression, we define two strains: the maximum site strain $\eps_1=b/(4\pi\rc)$, and the maximum strain at the cylinder edge $\eps_0=b/(4\pi R)$.  Then,
\be
\begin{split}
n(\eps) &= 2\disl\cdot\int_{\rc}^{\min\{R,b/4\pi\eps\}} dr\frac{r}{\left(\left(\frac{b}{4\pi r}\right)^2 - \eps^2\right)^{1/2}} \\
&= 2\disl\left(\frac{b}{4\pi}\right)^2\int_{\max\{\eps_0,\eps\}}^{\eps_1} dx\; x^{-3}(x^2-\eps^2)^{-1/2}\\
&= \frac{2\eps_0^2}{\pi}\int_{\max\{\eps_0,\eps\}}^{\eps_1} dx\; x^{-3}(x^2-\eps^2)^{-1/2}.
\end{split}
\ee
For $|\eps|>\eps_0$, this gives
\be
n(\eps) = \frac{1}{\pi}\left\{\left(\frac{\eps_0}{\eps_1}\right)^2\frac{\sqrt{\eps_1^2 - \eps^2}}{\eps^2} + \frac{\eps_0^2}{\eps^3}\arccos\left(\frac{\eps}{\eps_1}\right)\right\}
\ee
and for $|\eps|<\eps_0$, this gives
\be
n(\eps) = \frac{1}{\pi}\left\{\left(\frac{\eps_0}{\eps_1}\right)^2\frac{\sqrt{\eps_1^2 - \eps^2}}{\eps^2} - \frac{\sqrt{\eps_0^2-\eps^2}}{\eps^2} + \frac{\eps_0^2}{\eps^3}\left[\arcsin\left(\frac{\eps}{\eps_0}\right) - \arcsin\left(\frac{\eps}{\eps_1}\right)\right]\right\}
\ee
These two expressions can be written in terms of the ratio $\eta = \eps_0/\eps_1<1$ as
\be
n(\eps) =
\begin{cases}
\frac{1}{\pi\eps^{3}}\bigg( \eps\eta^2\sqrt{\eps_1^2 - \eps^2} +\eps_0^2\arccos\left(\eps/\eps_1\right) \bigg)
& : |\eps|>\eps_0\\
\frac{1}{\pi\eps^{3}}\bigg( \eps\eta^2\sqrt{\eps_1^2 - \eps^2} - \eps\sqrt{\eps_0^2 - \eps^2}
& : |\eps|<\eps_0\\
\ +\eps_0^2\left[\arcsin\left(\eps/\eps_0\right) -\arcsin\left(\eps/\eps_1\right)\right] \bigg)
\end{cases}
\label{eqn:dos}
\ee
The general scaling $n\sim |\eps|^{-3}$, similar to Kirchheim.\cite{Kirchheim1982}  If we integrate this density of states over all strains, we have
\be
\int_{-\eps_1}^{\eps_1} d\eps\;n(\eps) = 1-\eta^2
\ee
which accounts for the ``missing'' core states, which are a fraction $\eta^2=\rc^2/R^2$ of all possible sites.  We add back the core sites that make up $6\rc^2\disl$ of all possible sites; half have tensile strain $+\eps_1$, and the other half have compressive strain $-\eps_1$.  In our sample, the dislocation density is $\disl = 10^{11}\text{cm}^{-2}$, so $R=1/\sqrt{\pi\disl} = 63.7b = 19\text{nm}$, the maximum site strain is $\eps_1=b/(4\pi\rc)=0.05$, and the maximum strain at the cylinder edge is $\eps_0=b/(4\pi R)=1.25\EE{-3}$, with a ratio of $\eta = \eps_0/\eps_1 = 0.025$, and with a core occupancy of $6\rc^2\disl =2\cdot0.576\EE{-3}$.

The thermodynamics of hydrogen in Pd requires considering not just the site strain from a dislocation, but also from neighboring hydrogen atoms.  The site adjacent to a hydrogen interstitial in Pd experiences strain due to the occupancy of the hydrogen site; this strain, in term, affects the site energy.  In a 256-atom Pd supercell calculation of a hydrogen interstitial, the relaxation neighboring the hydrogen interstitial site is \textit{expanded} by $\Delta\eps=6.864\EE{-3}$; this produces a lowered site energy of approximately $\DE=-23\meV$.  It should be noted that this is purely classical approximation---it ignores not only electronic structure effects, but zero-point displacement of the two hydrogen atoms.  However, it should give the correct order of magnitude for the strength of interaction, and it suggests a propensity for ordering on the hydrogen sublattice.

To account for the weak H-H binding on the hydrogen distribution and site occupancy, we consider a simple self-consistent mean-field model.  A site with energy $E$ (or, alternately, strain $\eps$) will be shifted by $\DE$ if any of its neighbors are occupied, and unshifted if all are unoccupied.  We will ignore spatial variations in the local site occupancy, and so approximate the probability of each neighboring site being occupied with the site occupancy $\site$.  As there are twelve possible nearest-neighbor sites in the FCC hydrogen sublattice, the fraction of sites where all twelve neighbors are unoccupied is $(1-\site)^N$ with $N=12$; hence, each site now has two possible energy levels: a fraction $(1-\site)^N$ with energy $E$ and a fraction $1-(1-\site)^N$ with energy $E+\DE$.  To be in equilibrium, these sites have occupancies of $f_0 = (\exp(\beta(\mu-E))+1)^{-1}$ and $f_1=(\exp(\beta(\mu-E-\DE))+1)^{-1}$, respectively.  Thus, the occupancy of a site satisfies the self-consistent equation
\be
\site = f_1 + (1-\site)^N (f_0-f_1)
\label{eqn:occsc}
\ee
This equation is solved for $\site$ at each site given its energy $E$, and the chemical potential $\mu$; the occupancy is integrated over the density of sites to determine the total concentration of hydrogen.  \Eqn{occsc} can be solved approximately (to $10^{-4}$) by making a quadratic approximation around $\site\approx f_1$ to $\site = g(\site)$.  Defining the function and its first two derivatives at $f_1$,
\be
\begin{split}
g^{(0)} &= f_1 + (1-f_1)^N(f_0-f_1)\\
g^{(1)} &= -N(1-f_1)^{N-1}(f_0-f_1)\\
g^{(2)} &= N(N-1)(1-f_1)^{N-2}(f_0-f_1)
\end{split}
\ee
the quadratic approximate self-consistent solution is
\be
\site = 2\left[g^{(0)} - g^{(1)}f_1 + \frac12 g^{(2)} f_1^2\right]\cdot
\left[\left(1-g^{(1)}+g^{(2)}f_1\right) + \left( (1-g^{(1)})^2 + 2g^{(2)}(f_1-g^{(0)}) \right)^{1/2}\right]^{-1}
\ee
This self-consistent mean-field model accounts for the hydrogen-hydrogen attraction, and the primary effect is at low (but above zero) temperature where the ordering competes with entropy; it produces somewhat higher hydrogen occupancies than would be expected without any H-H interaction.  This approximate thermodynamic model is not accurate when the hydrogen occupancy becomes large; for example, it does not account for the formation of PdH$_{0.63}$ before the formation of PdH.

\bfig
\includegraphics[width=\smallfigwidth]{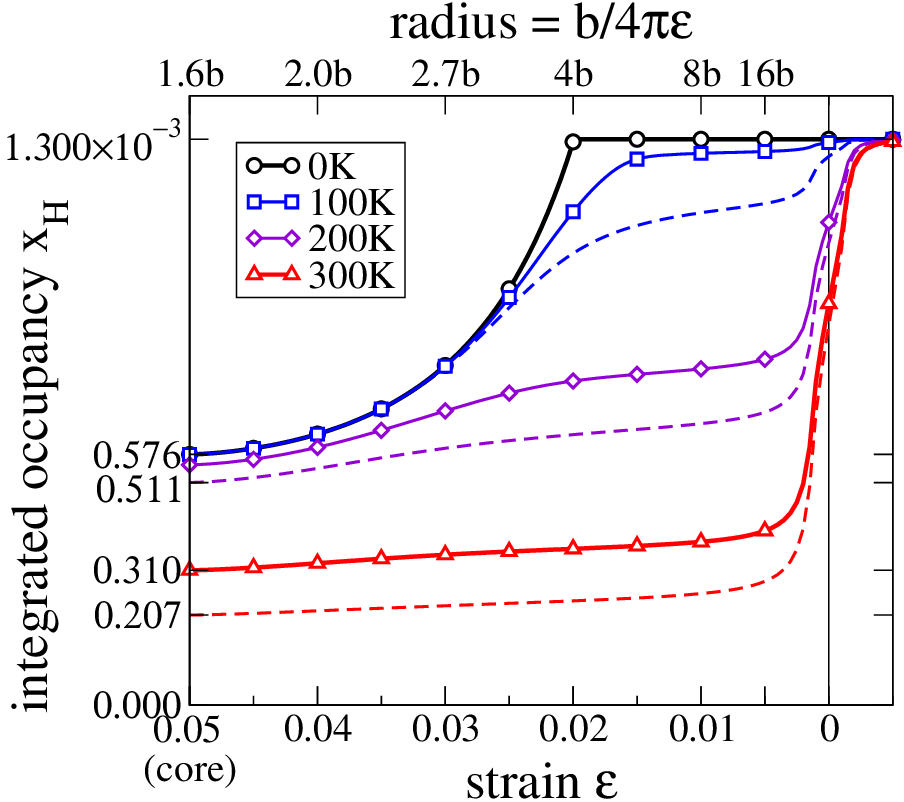}
\caption{(Color online) Integrated occupancy of hydrogen around dislocations in Pd with temperature for $\DE=23\meV$ (solid) and $\DE=0$ (dashed).  The integrated occupied density of sites goes from the most favored sites (dislocation cores) through the range of volumetric strain around the dislocation core; all hydrogen solutes are accounted for at the saturation concentration of $x_\text{H}=1.3\EE{-3}$.  The occupancy follows a Fermi function for $\DE=0$, and the effect of H-H coupling is to maintain the nanoscale hydride to slightly higher temperatures.  At 0K, the Cottrell atmosphere has a sharp boundary at $r=4b=7.9\text{\AA}$.  At 100K, the atmosphere shows only small spreading away from the core, while at 200K there is an increasing occupancy for H at 0 strain.  At 300K, the atmosphere is dissolving, with decreased occupancy in the core as well as around the dislocation.}
\label{fig:inter-occ}
\efig

\Fig{inter-occ} shows the formation of Cottrell atmosphere at low temperatures and dissolution near room temperature, including the difference between integrated occupancies assuming $\DE=0$ and $\DE=23\meV$.  Qualitatively, assuming $\DE=0$ shows similar behavior to $\DE=23\meV$, with dissolution of the nanoscale hydride between 200K and 300K.  The primary effect of the H-H binding is to maintain a slightly higher hydrogen concentration in the dislocation cores.  \Fig{inter-occ} shows the \textit{integration} of site occupancy, starting from the core; the derivative with strain gives the fraction of H at a specific strain.  At 0K and 100K the core is fully occupied; hence, the integrated occupancy starts at $0.576\EE{-3}$.  At 200K the core is 96\%\ occupied, falling to 54\%\ occupancy at 300K.  As temperature rises, lower strain sites have an increased occupancy due to entropy, and sites near the core are less populated---the ``dissolution'' of the Cottrell atmosphere, though the core still has hydrogen.  The fractional occupancy of sites near zero strain decays exponentially, but as the number of sites is growing as $|\eps|^{-3}$ most of the hydrogen is well dispersed at higher temperatures.

Prediction of vibrational excitations for hydrogen requires sampling of different Pd displacements neighboring the H atom to determine the potential energy.  Hydrogen is surrounded by 6 Pd neighbors at $\frac{a}{2}\langle100\rangle$.  These six neighbors are displaced according to the thermal occupation of phonons, including the quantum-mechanical zero-point motion.  The displacements provide an important broadening of the hydrogen vibrational excitation spectra, as the light hydrogen atom evolves in a Born-Oppenheimer-like manner (valid as $M_\text{H}\approx 10^{-2}M_\text{Pd}$), sampling the local potential energy from the neighboring Pd.  To compute a density of excitation energies for the H atom, we need to sample the possible displacements for neighboring atoms at a temperature $T$.  For the highest frequency excitation of Pd, 8THz ($\hbar\omega = 33\meV$), $x_0/\sqrt{2}=0.025\text{\AA}$; at 300K, $\bar x = 0.033\text{\AA}$.  The Gaussian distribution of displacements for a harmonic oscillator (see Appendix) provides the basis for random sampling displacements for Pd atoms from independent Gaussians of width $\bar x(\omega_n(\qv), T)$ for each phonon mode $\omega_n(\qv)$ in the Brillouin zone.  Let $\DR(\Rv)$ be the $3\x3$ force-constant matrix between an atom at 0 and $\Rv$; moreover, let $\uv(\Rv)$ be the displacement vector for an atom at $\Rv$.  Then, the Fourier transforms of $\DR$ and $\uv$ are
\be
\begin{split}
\Dk(\qv) &= \sum_{\Rv} \DR(\Rv) e^{i\qv\cdot\Rv}\\
\uk(\qv) &= \frac1{\sqrt N} \sum_{\Rv} \uv(\Rv) e^{i\qv\cdot\Rv}
\end{split}
\ee
for a bulk system of $N$ atoms.  The inverse Fourier transforms are
\be
\begin{split}
\DR(\Rv) &= \frac1N \sum_{\qv} \Dk(\qv) e^{-i\qv\cdot\Rv}\\
\uv(\Rv) &= \frac1{\sqrt N} \sum_{\Rv} \uk(\qv) e^{-i\qv\cdot\Rv}
\end{split}
\ee
where we have used the fact that there are also $N$ q-points in the Brillouin zone summation.  Note also that,
\be
\sum_{\Rv} \left|\uv(\Rv)\right|^2 = \sum_{\qv} \left|\uk(\qv)\right|^2.
\ee
Then, the displacements $\uk(\qv)$ can be written as the sum of three Gaussian distributed random variables $\alpha_n(\qv)$, multiplied by the corresponding width $\bar x(\omega_n(\qv), T)$ and normalized eigenvector of $\Dk(\qv)$, $\uv_n(\qv)$.  In reciprocal space, the sampled displacement $\uk(\qv)$ is
\be
\uk(\qv) = \sum_{n=1}^3 \alpha_n(\qv) \uv_n(\qv) \left[\frac{\hbar}{2m\omega_n(\qv)} \coth\left(\hbar\omega_n(\qv)/2\kB T\right)\right]^{1/2}
\ee
The final step is to inverse Fourier transform all of the displacements, and to remove the center-of-mass shift for the the six neighbors surrounding the H atom at $\{\rv\}$.  In the sum over the discrete $\qv$ in the Brillouin zone, the weight of each point $w(\qv) = 1/N$, so
\be
\Delta u(\rv) = \sum_{n\qv} \alpha_n(\qv) \uv_n(\qv) \left[\frac{w(\qv) \hbar}{2m\omega_n(\qv)} \coth\left(\hbar\omega_n(\qv)/2\kB T\right)\right]^{1/2}
\cdot\bigg\{\cos(\qv\cdot\rv) - \frac{1}{6}\sum_{\rv'} \cos(\qv\cdot\rv')\bigg\}
\ee
This requires $3N-3$ random Gaussian variables $\alpha_n(\qv)$ to produce one sample of displacements for Pd atoms neighboring the hydrogen atom at a temperature $T$.

The force-constants for Pd come from \abinit\ via a direct-force technique\cite{Kunc82} with a $4\x4\x4$ simple-cubic supercell; this reproduces the elastic constants and phonons within 5\%.  We use a discrete $16\x16\x16$ Monkhort-Pack mesh\cite{Monkhorst76} of $q$-points the Brillouin zone.  With 40,000 displacements for each temperature (0K to 300K), in the dislocation core and strains from +0.05 to --0.01 in 0.01 increments, we compute vibrational excitations for H in Pd.  Given the H potential energy, we solve the Schr\"odinger equation numerically.  For each Pd displaced environment, we find the minimum energy position for H, and expand the potential as a fourth-order polynomial in H displacement, and compute the three lowest-lying excitations using a Hermite-polynomial basis.\cite{Lawler2010}  This gives 120,000 excitation energies, binned into 1meV bins.  Thus, we predict vibrational density of states for H in a dislocation core, and at strains from +0.05 to --0.01 at 0K, 100K, 200K, and 300K.

To efficiently describe the energy landscape for a hydrogen atom in a variety of interstitial sites---including small displacements of Pd due to quantum-thermal vibrations---we optimize an embedded-atom method-like potential for H based on its distance to six neighboring sites.  The embedded-atom method\cite{Stott1980, Norskov1980, Puska1981, Daw93} can work well for describing the energy of atoms in metallic systems: neighboring atoms have overlapping charge densities at a site, and atoms experience an ``embedding energy'' due to that local environment.  As we are interested in describing H accurately for a small range of environments, we define a potential based on similar ideas, but make the fitting parameters as linear as possible so that overfitting can be easily identified, and good transferability achieved.  From previous calculations,\cite{Lawler2010} we have a large amount of force-displacement data for H in different environments (58 displacements in the dislocation core, 40 displacements in unstrained Pd, and 32 displacements in +5\%\ strained Pd).  This fitting database gives sufficient coverage that our potential will be used to \textit{interpolate} rather than \textit{extrapolate}.  The general form of the total energy in terms of the H-Pd distances $r_m$ is
\be
\begin{split}
E_\text{H}(\{r_m\}) &= \sum_{d=2}^D U_d \rho^d + \sum_m\left[\sum_{c=1}^C \phi_c r_m^c\right]\\
\text{where\quad} \rho &= \sum_m e^{-ar_m}
\end{split}
\ee
where $D$ and $C$ determine the polynomial order of the embedding energy $U(\rho)$ and the pair potential $\phi(r)$; besides the coefficients $U_d$ and $\phi_c$, there is the parameter $a$ which determines decay length of the density.  This means that the energies (and forces) are linear in all parameters except $a$; we can easily optimize the parameters by solving for $U_d$ and $\phi_c$ for a given $a$ with the smallest mean-squared error in the forces (weighted by the force magnitude).  Hence, for any choice of $D$ and $C$, we can find optimal parameters to accurately reproduce the DFT forces.  To optimize the choice of $D$ and $C$, we computed the leave-one-out cross-validation score (CVS) for each optimal set of parameters; $D=2$ and $C=5$ had the lowest CVS.  This fit ($E_\text{H}$ in eV, $r_m$ in \AA),
\be 
\begin{split}
E_\text{H}(\{r_m\}) =&\; 
4025.39 \Big(\sum_m e^{-3.4715 r_m}\Big)^2 \\
&+ \sum_m \Big\{ -131.94 r_m + 119.41 r_m^2
-54.073 r_m^3 + 12.1883 r_m^4 - 1.09167 r_m^5\Big\}
\end{split}
\label{eqn:EAM}
\ee
had no error larger than 10\%\ in any of the forces, and reproduced the H excitation spectra of the direct DFT calculation to within 2meV.  As $r\lesssim 2\text{\AA}$, the contribution of the higher order polynomial coefficients is decreasing to larger orders.

\section{Results}
\bfig
\includegraphics[width=\smallfigwidth]{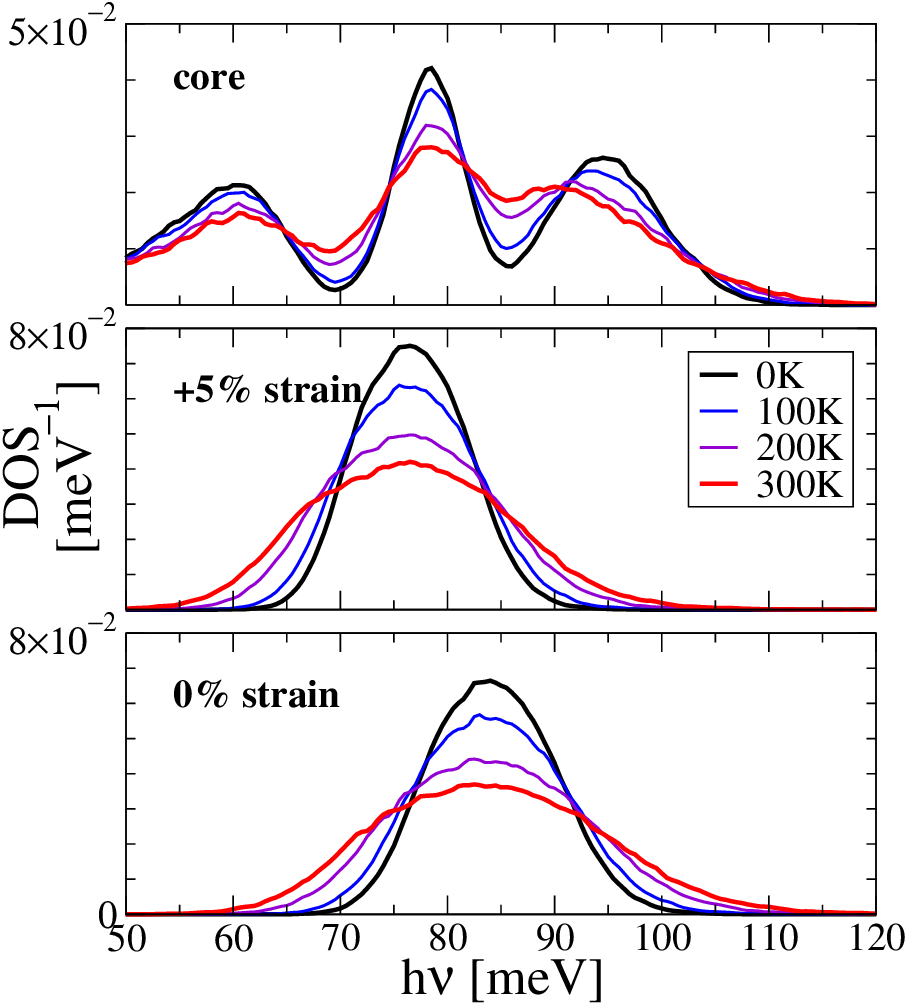}
\caption{(Color online) Calculated vibrational density of states for hydrogen in Pd with temperature.  Increasing temperature produces larger displacements of Pd beyond the zero-point motion at 0K; this increases the spread in the vibrational excitations.  The central peaks for the three sites are temperature independent.  Peak broadening smears the low and high excitations in the dislocation core at room temperature.}
\label{fig:DOS-sites}
\efig

\Fig{DOS-sites} shows the predicted vibrational density of states for hydrogen at equilibrium zero strain, a 5\%\ expanded site, and in the partial core.  Increasing temperature broadens the excitation spectra with increased vibration of neighboring Pd atoms.  There is no shift in the peak position with temperature due to Pd vibration, but only from strains.  The dislocation core environment breaks cubic symmetry, giving three peaks below 120meV.\cite{Lawler2010}  Temperature widens the peaks above and below 78meV on each side of the central peak at room temperature.  Hence, despite dislocation core occupancy at room temperature, it is difficult to experimentally identify H in the dislocation core except at low temperatures.

\bfig
\includegraphics[width=\smallfigwidth]{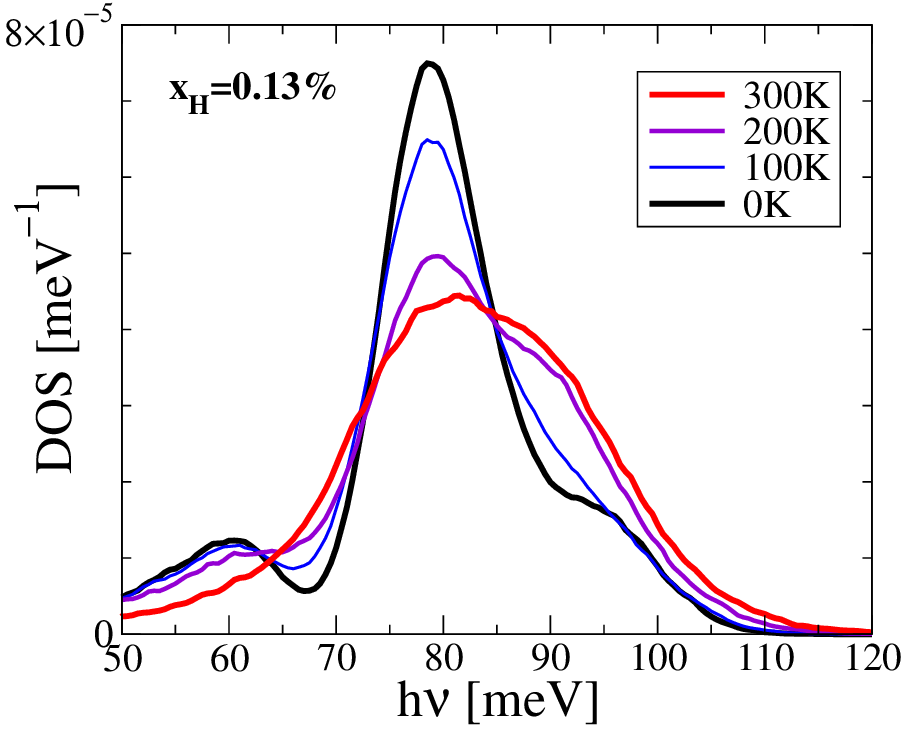}\\
\includegraphics[width=\smallfigwidth]{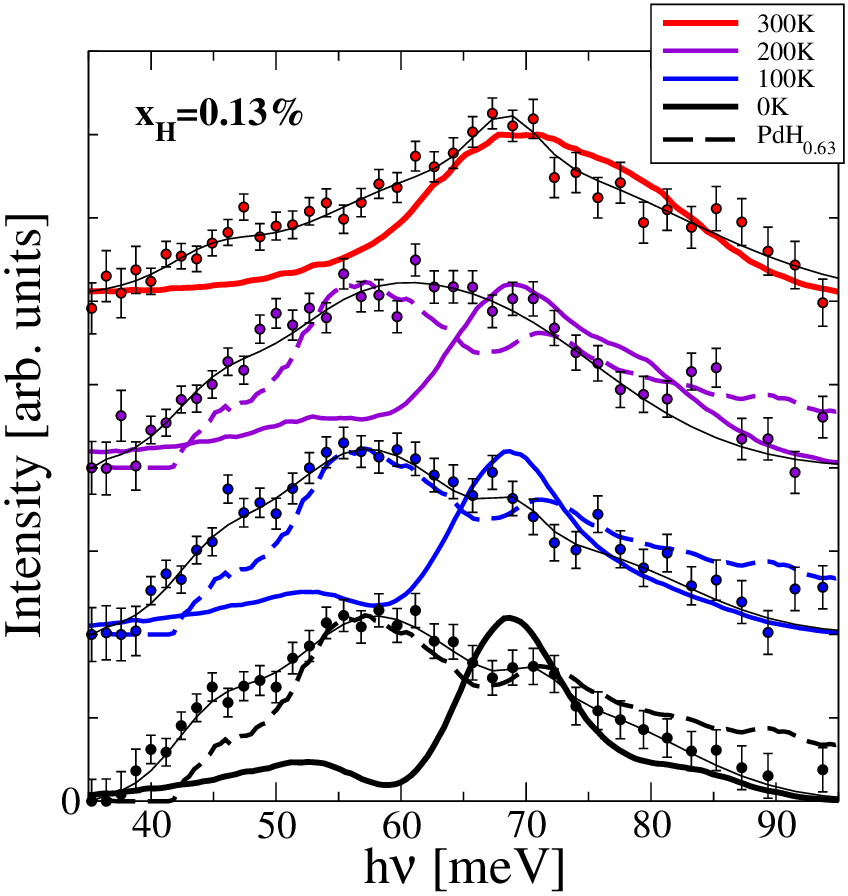}
\caption{(Color online) The predicted vibrational density of states and inelastic neutron scattering intensity for 0.13at.\% H in Pd as a function of temperature.  The temperature determines both the occupancy of states for H (c.f. \Fig{inter-occ}) and the  vibrational spectra for all states (c.f. \Fig{DOS-sites}); taken together, we predict the density of states in the top figure.  To compare with IINS measurements, we scale intensity by $1/\sqrt{h\nu}$, equalize amplitudes, and scale energy by 7/8 (DFT/experimental discrepancy).  The agreement in line shape at 300K confirms that the main cause of peak broadening is Pd vibration.  At lower temperatures, the formation of a Cottrell atmosphere creates nanoscale regions with high hydrogen concentration.  The scattering signal from $\beta$-PdH has a width similar to the experimentally measured spectrum at 0K;\cite{Heuser2008} the difference from the \abinit\ prediction is due to the dispersion of a hydride which is missing in our calculation of isolated hydrogen vibrations.  The computed PdH$_{0.63}$ spectra (dashed lines) has dispersion but is a harmonic approximation for hydrogen.  The signal change can estimate the fraction of nanoscale hydrides at dislocation cores.}
\label{fig:DOS-predict}
\efig

\Fig{DOS-predict} shows the predicted vibrational spectra for 0.13at.\% H in Pd as a function of temperature, and the comparison with inelastic neutron scattering measurements.  Combining the site-occupancy data from \Fig{inter-occ} with the predicted vibrational spectra in \Fig{DOS-sites}, we predict the expected measured vibrational spectra with temperature.  To compare with the experimental measurements, we scale all of our peak heights to be equal, scale intensity by $1/\sqrt{h\nu}$ to produce a scattering cross-section under the condition of variable incident energy and fixed final energy (as is the case for the measured IINS spectra reported here), and scale energy by 7/8.  The latter scaling corresponds to a needed softening of the DFT calculations of vibrational spectra for H in Pd compared with experimental measurements; the overestimation of vibrational excitation is independent of exchange-correlation potential and treatment of H and Pd ionic cores\cite{Lawler2010} and is consistent with earlier fully-anharmonic calculations of isolated hydrogen in Pd.\cite{Elsasser1991}  The experimentally measured line shape is in good agreement with the prediction of scattering at room temperature, but the shapes begin to deviate as temperature is lowered.

Lowering temperature forms a Cottrell atmosphere and the predicted scattering cross-section shifts and narrows; the shift in peak energy agrees with the experimental measurements, but the peak narrowing does not.  At 300K, hydrogen is primarily in low strain environments, and has a peak widened primarily by vibration of Pd neighbors.  As temperature is lowered, \abinit\ calculations predicts a shift of the peak to lower frequencies as higher strain sites and the dislocation core is preferentially occupied; this matches the experimental measurement as well.  However, the \abinit\ calculations predict a narrowing of spectra; this narrowing is due to the smaller displacements of Pd neighbors producing less random distortion of the potential energy.  As the Cottrell atmosphere forms, the local hydrogen concentration near the dislocation core is very high, forming hydride phases in nanoscale cylinders.  This corresponds well with recent small-angle neutron scattering measurements at low temperatures and hydrogen concentrations in deformed single-crystal Pd.\cite{Heuser2011}  The vibrational spectrum of $\beta$-PdH is wider due to H-H interactions;\cite{Heuser2008} this dispersion is lacking in the \abinit\ calculations due to the difficulty of predicting fully anharmonic dispersion relations.  We have computed the harmonic bulk PdH$_{0.63}$ vibrational density-of-states that includes dispersion, but lacks anharmonicity; the comparison with the IINS signal from 0K to 200K strongly backs up the presence of hydride.  Fitting the experimental intensity to a linear combination of the two predicted intensities suggests all hydrogen is in hydride and none is free at 0K; a 9:1 ratio at 100K; a 9:4 ratio at 200K; and dissolution of the hydride at 300K.  Hence, we conclude that the Cottrell atmosphere is forming of nanoscale hydride particles near dislocation cores, despite the low total hydrogen concentration in the sample, to explain the changes in vibrational spectra.

\section{Conclusion}
Combining the experimental measurement of hydrogen vibrational spectra with \abinit\ calculations of vibrational spectra with temperature, we can identify the formation of Cottrell atmosphere leading to nanoscale hydride precipitates at dislocation cores.  By separating the sources of spectral broadening---dispersion in hydrides at low temperatures, and thermal broadening from Pd vibration of neighbors---and the causes of a peak shift, we have \textit{in situ} characterization of the hydrogen environment evolution with temperature.

\begin{acknowledgments}
This research was supported by NSF under grant number DMR-0804810, and in part by the NSF through TeraGrid resources provided by NCSA and TACC.  We acknowledge the support of the National Institute of Standards and Technology, U.S. Department of Commerce, in providing the neutron research facilities used in this work.
\end{acknowledgments}

\appendix
\section{Harmonic displacement distribution at finite temperature}
For an isolated harmonic oscillator of mass $M$ and natural frequency $\omega$, we want to determine the probability distribution of displacements $x$ from equilibrium.  The state energies are $E_n = \hbar\omega(n+1/2)$, and so the probability of being in state $n$ at temperature $T$ ($\beta=(\kB T)^{-1}$) is
\be
z_n = \frac{e^{-\beta E_n}}{\sum_m e^{-\beta E_m}} = e^{-n\beta\hbar\omega}\left(1-e^{-\beta\hbar\omega}\right)
\ee
The wavefunctions are
\be
\psi_n(x) = (2^n n!)^{-1/2} (\pi x_0^2)^{-1/4}e^{-x^2/2x_0^2} H_n(x/x_0)
\ee
for natural length $x_0 =\sqrt{\hbar/m\omega}$, and Hermite polynomial $H_n$.  Then the probability distribution of displacement $x$ is
\be
\begin{split}
P(x) &= \sum_{n=0}^\infty z_n \left|\psi_n(x)\right|^2\\
&= (1-e^{-\beta\hbar\omega})\sum_{n=0}^\infty \frac{e^{-n\beta\hbar\omega}}{2^n n! \sqrt\pi x_0} e^{-x^2/x_0^2} H_n^2(x/x_0)\\
&= \frac{1-e^{-\beta\hbar\omega}}{\sqrt{1-e^{-2\beta\hbar\omega}}}\frac{1}{\sqrt\pi x_0} \exp\left(-\frac{1-e^{-\beta\hbar\omega}}{1+e^{-\beta\hbar\omega}}\cdot\frac{x^2}{x_0^2}\right)\\
&= \frac{1}{\sqrt{2\pi} \bar x(\omega,T)}\exp\left(-\frac{x^2}{2\bar x^2(\omega,T)}\right)
\end{split}
\ee
where
\be
\bar x(\omega,T) = \left(\frac{\hbar}{2m\omega} \coth\left(\frac{\beta\hbar\omega}{2}\right)\right)^{1/2}
\ee
is the thermal Gaussian width; the simplification is possible by using Mehler's Hermite polynomial formula,\cite{Mehler1866,Watson1933}
\beu
\sum_{n=0}^\infty \frac{H_n(x)H_n(y)}{n!}\left(\frac{w}{2}\right)^n = (1-w^2)^{-1/2} \exp\left[\frac{2 x y w - \left(x^2+y^2\right)w^2}{1-w^2}\right].
\eeu
In the low temperature limit, $\bar x\approx x_0/\sqrt{2}$ as expected from zero-point motion; and in the high temperature limit, $\bar x\approx (\kB T/m\omega^2)^{1/2}$, as expected from the equipartition theorem.

\end{document}